\documentclass[preprint]{aastex} 
\usepackage{graphicx}
\usepackage{hyperref}
\usepackage{pdfpages}

\newcommand{\src}{1E~1048.1$-$5937}

\begin{document} 

\title{Mapping the  Surface  of  the  Magnetar  1E~1048.1$-$5937  in  Outburst  and
  Quiescence Through Phase Resolved X-ray Spectroscopy}
 
\author{Tolga G\"uver$^{1}$, Ersin G\"o\u{g}\"u\c{s}$^{2}$ \& Feryal \"Ozel$^{3}$}

\affil{$^{1}$ Istanbul University, Science Faculty, Department of
Astronomy and Space Sciences, Beyaz{\i}t, 34119, Istanbul, Turkey}

\affil{$^{2}$ Sabanc\i~University, Faculty of Engineering and Natural 
  Sciences, Orhanl\i ~Tuzla 34956 Istanbul Turkey}

\affil{$^{3}$  Department of Astronomy,  University of  Arizona, 933
    N. Cherry Ave., Tucson, AZ 85721}

\begin{abstract} 
 
We model the pulse profiles and the phase resolved spectra of the
anomalous X-ray pulsar \src\ obtained with XMM-Newton to map its
surface temperature distribution during an active and a quiescent epoch.
We develop and apply a model that takes into account the relevant
physical and geometrical effects on the neutron star surface,
magnetosphere, and spacetime. Using this model, we determine the
observables at infinity as a function of pulse phase for different
numbers and sizes of hot spots on the surface. We show that the pulse
profiles extracted from both observations can be modeled with a single
hot spot and an antipodal cool component. The size of the hot spot
changes from $\approx 80^{\circ}$ in 2007, 3 months after the onset of a dramatic flux
increase, to $\approx 30^{\circ}$ during the quiescent observation in 2011, when the
pulsed fraction returned to the pre-outburst $\approx$~65\% level. For the 2007
observation, we also find that a model consisting of a single 0.4~keV
hot spot with a magnetic field strength of $1.8~\times~10^{14}$~G accounts for the
spectra obtained at three different pulse phases but underpredicts the
flux at the pulse minimum, where the contribution to the emission from
the cooler component is non-negligible. The inferred temperature of the
spot stays approximately constant between different pulse phases, in
agreement with a uniform temperature, single hot spot model. These
results suggest that the emitting area grows significantly during
outbursts but returns to its persistent and significantly smaller size
within a few year timescale. \end{abstract} 
  

\section{Introduction}

Multiwavelength  observations and  monitoring  campaigns  in the  past
decade  revealed  a large  number  of  fundamental characteristics  of
Anomalous X-ray Pulsars (AXPs)  and Soft-Gamma Repeaters (SGRs). There
is strong evidence  in favor of very high neutron  star magnetic field
strengths,  obtained from  observations  of their  rates of  spindown,
their  ultra-energetic   bursts,  and  their  high   persistent  X-ray
luminosities (see,  e.g., Woods \&  Thompson 2006 and  Mereghetti 2008
for detailed reviews). Indeed, the  magnetar model (Thompson \& Duncan
1995), in which a magnetic field in excess of $B \sim {\rm few} \times
10^{13}$~G  is responsible  for powering  the persistent  emission and
giving  rise  to  the  bursting   activity,  has  been  successful  in
explaining a wide variety of broad properties of these sources\footnote{See the  online AXP/SGR   
                       catalog                          at
 \href{http://www.physics.mcgill.ca/~pulsar/magnetar/main.html}{http://
 www.physics.mcgill.ca/~pulsar/magnetar/main.html} for a summary  of
 their properties. (Olausen \& Kaspi 2014)}.
 
Despite   the  significant   progress  in   understanding  the   basic
characteristics of  AXPs and SGRs,  a number of  outstanding questions
remain that are related to the properties of their magnetic fields and
of their  long-term emission.   For example,  AXPs and  SGRs generally
exhibit  high  spindown rates,  which  indicate  high inferred  dipole
magnetic  fields  for their  typical  spin  periods  in the  range  of
2$-$12~s.   However, recently  a  number of  radio  pulsars with  high
spindown  rates   that  do  not   behave  like  magnetars   have  been
discovered. Conversely, some  AXPs that possess low  rates of spindown
but show  magnetar-like behavior have  also been observed  (see, e.g.,
Rea  et  al.\ 2010,  2012, 2014; Zhou et al. 2014).   Even 
though the  magnetic  field geometry is thought  to play a role in
shaping  the broadband emission characteristics of AXPs  and SGRs (e.g.,
G\"uver et  al.\ 2011; Tiengo et  al.\ 2013),  there  have  been no 
direct  probes  of their  field strength and its variation at the
surface.

Second, these sources  show a variety of transient  phenomena, such as
bursts, flares,  outbursts, as  well as  significant changes  in their
persistent fluxes (see,  e.g., Kaspi 2007; Rea \&  Esposito 2011).  It
is unclear what  the different triggers for these  events are, whether
the  same triggers  recur, and  what properties  of the  neutron stars
change  during  the  transient  events.    For  example,  in  one
  interpretation, a  change in  the magnetospheric twist  angle causes
  the  outbursts,  with  accompanying spectral  changes  (Thompson  et
  al.\ 2007). In a different model, \"Ozel \& G\"uver (2007) attribute
  the  outbursts and  post-burst flux  enhancements to  purely thermal
  changes   in  the   surface,   without  any   modification  to   the
  magnetosphere.

Third,  the  observed pulsed  emission  exhibits  a variety  of  pulse
morphologies and  amplitudes for  different sources (e.g.,  Gavriil \&
Kaspi 2002).   While the  gross characteristics of  these morphologies
can  be understood  within  the  context of  surface  emission from  a
magnetar (e.g., \"Ozel  2002), pulse-phase-dependent observables, such
as the  size of  the emitting  region, its  inclination angle,  or the
compactness of the  neutron star, have not been studied  in detail and
in connection with the global neutron star properties.
 
One  approach toward  addressing  these questions  is  by mapping  the
strength of the magnetic field and the temperature distribution at the
neutron star  surface through pulse-phase resolved  spectroscopy, 
  which  is a  technique that  allows tracking  spectral changes  as a
  function of the rotational phase. Mapping these two parameters can,
in  principle, reveal  the number,  size,  and location  of the  X-ray
emitting   regions   and   help    understand   the   observed   pulse
shapes. Furthermore, obtaining the distribution of these parameters on
the  stellar  surface,  following  their  evolution  during  transient
events, and  comparing them between different  transient phenomena may
also provide better  insight into the nature of these  sources and the
underlying  physical mechanisms  for these  events (see,  e.g., An  et
al.\  2013;  Scholz  \&  Kaspi  2011; Ng  et  al.\  2011;  G\"uver  et
al.\ 2007).

 In  this paper,  we analyze the  pulse-phase resolved  spectra of \src\
 obtained  with XMM-Newton  in 2007  and 2011,  when it  was in outburst
 and in quiescence, respectively. \src\ is a good target for such a
 study because of its  highly active history and  its fairly
 high pulsed fraction, which has been observed to show significant
 variations over time.    \src\  was   discovered  by   Seward  et al.\
 (1982)  with the {\it Einstein}  satellite and is located  at a
 distance of $\approx$~9~kpc (Durant \& van Kerwijk 2006). Its lowest
 observed     quiescent     flux      is     $\sim     4-5     \times
 10^{-12}~\rm{erg}~\rm{cm^{-2}}~\rm{s^{-1}}$ (see,  e.g., Oosterbroek et
 al.\  1998;  Tam  et  al.\  2008;  and  the  discussion  below).
 Observations with {\it EXOSAT} revealed  a spin period of 6.44~s and a
 pulsed fraction, i.e., the ratio  of pulsed flux to total flux, of 65\%
 (Seward  et al.\  1986).  This  pulsed fraction  is one  of the largest
 among  all known  AXPs  and  SGRs,  indicating  a high degree of
 localization of the hot region on the surface, which produces the
 observed pulsed soft X-ray emission as the star rotates.

 \src\  has indeed been one  of the most active  and variable AXPs
  since  its discovery,  showing  bursts,  outbursts, and  significant
  changes in its  quiescent flux. Oosterbroek et  al.  (1998) reported
  factor of  4 flux variations over  a time span of  two decades.  In
the  Rossi  X-ray Timing  Explorer  (RXTE)  era, \src\  was  monitored
regularly (see, e.g., Kaspi et al.\ 2001; Dib, Kaspi \& Gavriil 2009).
This monitoring campaign  resulted in the detection  of SGR-like short
bursts from an AXP for the first time (Gavriil, Kaspi, \& Woods 2002).
Accompanying these bursts,  \src\ also showed outbursts  in its pulsed
flux in 2003  and 2004 (Gavriil \& Kaspi 2004;  Gavriil et al.\ 2006).
Finally,  it became  active again  in  2007, showing  a large  spin-up
glitch,  which is a sudden increase in its rotation frequency (Dib
  et  al.   2009).   X-ray   observations  with  imaging  instruments
revealed that, while  the total flux was seven  times greater compared
to the quiescent  levels, the pulsed fraction  decreased from $\approx
70$\% to  $\approx 30$\%  (Tam et  al.  2008).   This anti-correlation
between  the total  flux and  the  pulsed fraction  was also  reported
earlier for this source (see, e.g., Tiengo et al.\ 2005).

The first phase-resolved  spectral analysis of \src\  was performed by
Oosterbroek et al.\ (1998)   using BeppoSAX data; however, due to
the low  count rates, they  were unable to detect any significant
phase  dependence   in  the  spectral  parameters.    Later,  using  a
relatively short ($\approx~8$~ks)  XMM-Newton observation performed in
2000,  Tiengo et  al.\ (2002)  reported that  the spectral  parameters
obtained from phase integrated analysis can describe most of the phase
resolved data satisfactorily, with the only variation seen in the 
  overall normalization,  which is  proportional to the  source flux.
They  found a  significant spectral  variation only  during the  phase
interval corresponding  to the  minimum flux.  Pulse  profile modeling
and pulse-phase resolved spectroscopy of other AXPs and SGRs have also
been  performed  in the  past  using  phenomenological models  of  the
surface and magnetospheric  emission (see, e.g., Albano  et al.\ 2010;
Bernardini et  al.\ 2011).    In particular, Albano et  al. (2010)
  analyzed X-ray  data of XTE~J1810$-$197  and CXOU~J164710.2$-$455216
  using  a model  that  assumes a  globally  twisted magnetosphere  as
  presented by Nobili, Turolla, \& Zane (2008).  Their results suggest
  that during  outbursts, a  limited fraction  of the  stellar surface
  that  is  close  to  the  magnetic poles  is  heated  and  that  the
  subsequent spectral evolution is due  to the cooling and the changes
  in the size of this region.

In this paper,  we extend these earlier studies  and model pulse-phase
resolved  spectra of  \src\ using  a physical  model that  accounts for
radiative  processes in  a strong  magnetic field  for the  persistent
emission from a magnetar. In  this model, we determine the observables
as a  function of  pulse phase  by calculating  the emission  from the
magnetic neutron  star atmosphere, transporting this  emission through
its magnetosphere,  and accounting for the  general relativistic light
bending the  photons experience in  the strong gravitational  field as
they travel toward  a distant observer.  In this first  study, we work
within  the  framework of  a  simple  temperature and  magnetic  field
configuration on the stellar surface,  where the emission comes from a
single or  two antipodal  hot regions of  uniform temperature  $T$ and
uniform magnetic  field $B$.    When  modeling pulse  profiles, we
  also take into account the emission from an antipodal cool component
  originating   from  the   rest   of  the   surface.   However,   for
  phase-resolved  spectroscopy,  we focus  on  the  properties of  the
  active  hot  spot  and  neglect   the  contribution  from  this  dim
  component.

In  Section~\ref{modelsec}, we  discuss  the details  of the  spectral
model. In  Section~\ref{obsdata} we  describe the  XMM data,  while in
Section~\ref{sec_res}  we present  the  results from  the analysis  of
pulse  profiles  and the  phase  resolved  spectroscopy.  Finally,  in
Section~\ref{disc}, we discuss the implications of our results.
 
\section{Phase-Resolved Spectral Models of Magnetar Emission} 
\label{modelsec} 
 
The  observed energy  and pulse  phase distribution  of X-ray  photons
emitted from  a magnetar  are shaped  by a  number of  processes (see,
e.g., \"Ozel 2013).  Photons propagate  outward from the deeper layers
of the  stellar crust through  an atmosphere in  radiative equilibrium
characterized by  an effective  temperature $T$  and a  magnetic field
strength $B$.   In the case  of magnetars, the atmosphere  is strongly
magnetic ($> 10^{13}$~G) and is  expected to consist of a nearly
fully ionized  electron-proton plasma (Zavlin, Pavlov,  Shibanov 1996;
\"Ozel 2001; Ho  \& Lai 2001; see  also Ho et al.   2003 for partially
ionized cases).   The strong magnetic  field affects the  transport of
radiation in this plasma significantly:  it distorts the spectrum away
from  a blackbody  and causes  absorption like  features in  the X-ray
spectra due  either to ion cyclotron  lines or to the  polarization of
the  vacuum  itself  (\"Ozel  2001,   2003;  Ho  \&  Lai  2003).   The
polarization-dependent  transport in  the  local  magnetic field  also
determines the beaming of the radiation emerging from the surface.

To model the spectrum and beaming of the surface emission, we use the
atmosphere calculations performed by \"Ozel (2001, 2003) at different
effective temperatures and magnetic field strengths, assuming a plane
parallel atmosphere and a magnetic field vector that is locally
perpendicular to the surface.  This assumption about the magnetic
field geometry is appropriate for emission originating from a highly
localized region on the stellar surface, such as a magnetic pole, and
allows us to avoid the immense additional computational cost that is
introduced by the additional angle that would otherwise have been
required to specify the local direction of the magnetic field, and the
second additional angle we would need to consider in the radiative
transfer equation when the azimuthal symmetry around the surface
normal is broken in this configuration.  When the emitting region is a
more substantial fraction of the surface, the magnetic field may be
inclined with respect to the surface normal  and vary in
  magnitude at different latitudes.  Model atmosphere calculations
for inclined magnetic fields show that, for larger field inclinations,
the spectra emerging from the surface are somewhat softer at photon
energies significantly above the peak of the spectrum (see Figure~10
of Lloyd 2003).  The model spectra at these higher energies, however,
are dominated by resonant scattering in the magnetar magnetosphere
(see below), and therefore, ignoring the effects of the inclined
magnetic fields does not introduce a large correction to the models.

The interaction of the photons emitted from the surface with the
charged particles in the magnetosphere further shapes the observed
spectrum. In magnetars, the magnetosphere is expected to be populated
by energetic charged particles with a density that exceeds the typical
Goldreich-Julian density by many orders of magnitude (Thompson et
al.\ 2002).  The energy distribution of surface photons, therefore,
gets modified through resonant cyclotron scattering processes off of
these charges (e.g., Lyutikov \& Gavriil 2006; Fernandez \& Thompson
2007) as they pass through the magnetosphere. We take into account the scattering of the thermal photons from the
surface in the magnetosphere using the Schwarzschild-Schuster method for
solving the radiative transfer equation and angle averaged scattering
cross sections, as in the method followed by Lyutikov \& Gavriil (2006;
see Fernandez \& Thompson 2007 for a fully 3D treatment). In this
setup, the effect of the magnetospheric scattering on the spectrum is
captured by means of two parameters: the total scattering optical depth
$\tau$ and the average velocity of the electrons in the magnetosphere
$\beta$, where the latter is defined in units of the speed of light.
 
\begin{figure} 
  \centering 
\includegraphics[scale=0.4]{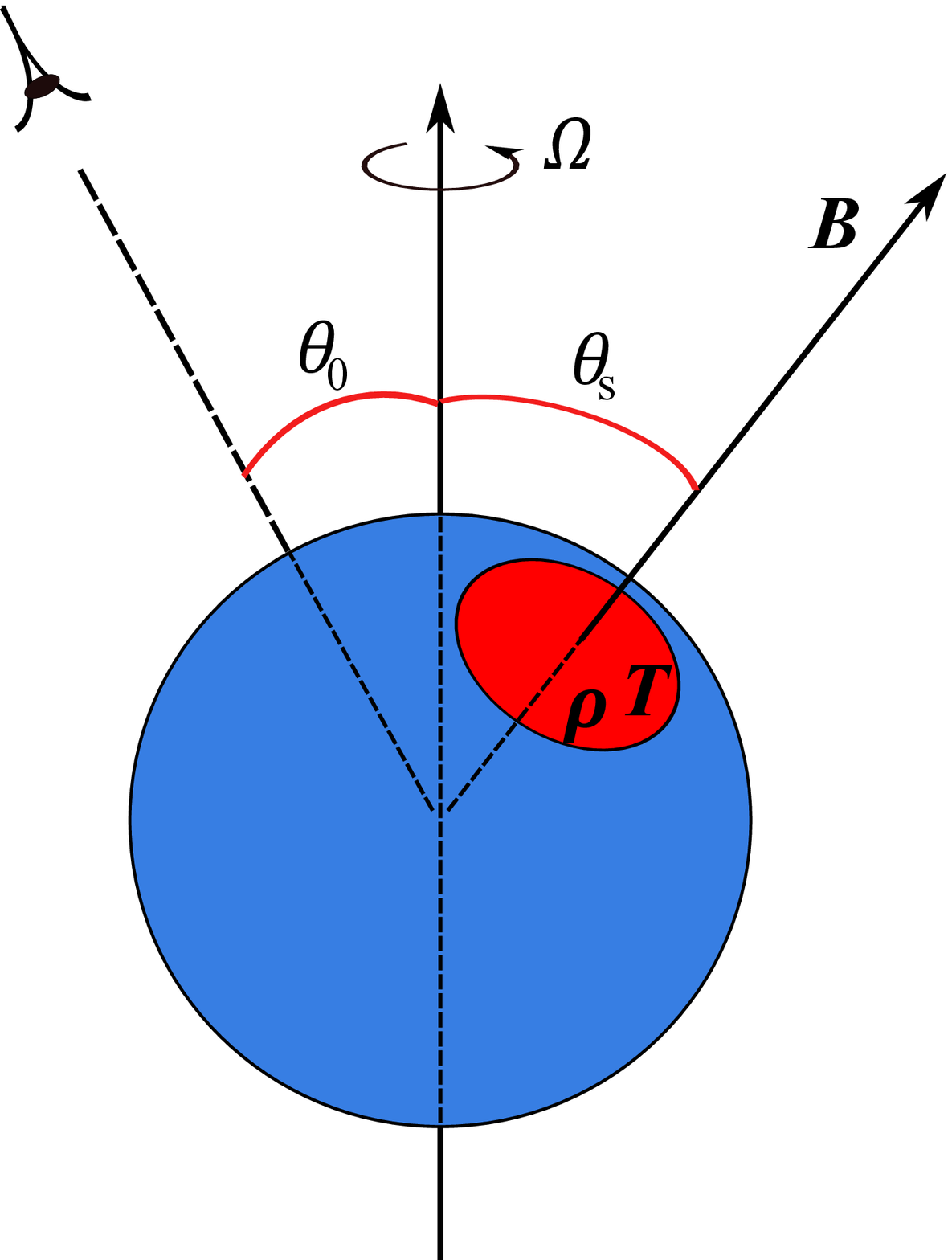} 
\caption{A schematic view of a neutron star with one emitting hot spot
  on the surface. The angle between  the rotation and magnetic axis is
  shown with $\theta_{\rm s}$ and  the angle between the rotation axis
  and the  observer is  shown with $\theta_{0}$.    $T$,  $B$, and
    $\rho$  denote the  temperature, magnetic  field strength  and the
    size of the spot, respectively.}
\label{model} 
\end{figure}

Finally, we take into account the effects of gravitational lensing and
redshift to  calculate the observables  at infinity. Because  AXPs and
SGRs spin very slowly, we use the Schwarzschild metric to describe the
neutron star spacetime and follow the method described in Pechenick et
al.\ (1983) to  compute spectra at different pulse  phases (see \"Ozel
2002  for a  detailed  description of  the algorithm).   Gravitational
lensing significantly alters the  observed pulsed fractions and causes
them to  be smaller compared  to the  Newtonian case, since  it causes
more of the stellar surface to  be observable (see, e.g., Pechenick et
al.\ 1983;  Psaltis et al.\  2000).  As  in \"Ozel\ (2002),  we assume
that there is either  a single hot spot or two  antipodal hot spots of
the same  size that are described  by a uniform temperature.   We also
assume that the rest of the neutron star does not radiate in the X-ray
band. Thus,  in this setup,  three geometric parameters  determine the
flux and spectrum  of photons that are observed as  a function of spin
phase, in addition to the neutron star compactness $GM/Rc^2$ (see also
Zavlin \&  Pavlov 2002).  As  shown in Figure \ref{model},  these are:
{\it (i)}  the sizes of  the hot spots emitting  in the X-rays  on the
surface of the  neutron star $\rho$, {\it (ii)} the  colatitude of the
center  of  the hot  spot  with  respect  to  the stellar  spin  axis,
$\theta_{\rm  s}$, and  {\it  (iii)} the  observer's inclination  with
respect to the stellar spin axis, $\theta_0$.
 
Accounting for the  physics of the atmosphere,  the magnetosphere, and
the gravitational lensing,  even in a simplified  geometry, requires a
total  of  eight  parameters   ($B$,  $T$,  $\tau$,  $\beta$,  $\rho$,
$\theta_{\rm  s}$,  $\theta_0$,  $GM/Rc^2$),  as  defined  above.   In
principle, models that incorporate all  of these effects should be fit
simultaneously to  the observed data  to constrain the  magnetic field
strength and  temperature distribution on  the surface of  a magnetar.
In reality,  however, sampling this  parameter space even with  a very
coarse resolution  that corresponds  to 10  values for  each parameter
results in $10^8$ models, each  of which yields the observed intensity
at 64  values of  the photon  energy (for  a relatively  coarse energy
resolution)  and  at  each of  the  30  pulse  phases  we use  in  the
calculation.   In order  to reduce  the computational  requirements of
modeling the pulse-phase resolved spectra  of magnetars, we follow the
approach outlined below.
 
For slowly spinning  neutron stars, there is a near  degeneracy in the
predicted pulse profiles between changing  the compactness of the star
and the size of the hot spot: the same pulse amplitude can be achieved
with a large compactness and a  small spot size or a small compactness
and a large spot size. In the analyses presented here, we fix the
  compactness by  assuming representative values for  the neutron star
  mass, M=1.4 M$_{\odot}$, and radius, $R=10$~km.
 
We then  use the  extraordinarily large  pulse amplitude  exhibited by
\src\ at some epochs to  place constraints on the geometric parameters
of  the model,  which are  not expected  to evolve  significantly with
time. As discussed in \"Ozel et  al.\ (2001), an amplitude as large as
75\% can be achieved with a single hot spot and only if one of the two
angles  $\theta_{\rm  s}$  or   $\theta_0$  is  close  to  $90^\circ$.
Moreover,  at these  small spin  frequencies, the  pulse amplitude  is
primarily  determined  by  the   product  $\sin  \theta_{\rm  s}  \sin
\theta_0$  (Poutanen \&  Beloborodov  2006), leading  to a  degeneracy
between these two  parameters. For this reason, we fix  one of the two
angles, $\theta_{\rm  s}$, to be  $90^\circ$. We will  further discuss
the validity of this assumption in Section~\ref{timingsec}.

Having fixed the  compactness of the neutron star  and having obtained
an estimate of some of the  geometric parameters of the model, we then
produce  a fine  grid  of  theoretical models  over  a  wide range  of
effective temperatures,  magnetic field strengths, as  well as optical
depths and the average velocities of the charges in the magnetosphere.
This computation resulted in an XSPEC compatible additive table model,
which is a version of  the Surface Thermal Emission and Magnetospheric
Scattering model (STEMS, G\"uver et  al. 2007, 2008), where the proper
geometric and beaming effects that  are necessary for a phase resolved
spectral  analysis are  taken  into  account. We  used  this model  to
perform a formal $\chi^2$ fitting  of the phase-resolved X-ray spectra
in Section~\ref{spec_res}.
 
\section{Observations \& Data Analysis} 
\label{obsdata} 
 
\src\ has been observed seven times with XMM-Newton. In this paper, we
focus  on  the  observations  that  were  long  enough  to  accumulate
sufficient number of source counts for phase resolved spectroscopy and
that took place  at epochs separated by a sufficiently  long period of
time to  allow us to  track its  long-term flux evolution  following a
transient event. There are two  such observations in the archive.  The
first  observation was  roughly  within three  months  after a  timing
glitch event observed in 2007 and  two months after the detection of a
short SGR-like burst. A simple  blackbody plus a power-law fit to the
X-ray  spectrum  extracted  from  this  observation  yields  an
unabsorbed   bolometric   flux   $    F   =   (2.75\pm0.02)   \times
10^{-11}~\rm{erg}~\rm{s}^{-1}~\rm{cm}^{-2}$,   in   the   2$-$10~keV
range.  The  second observation was  in 2011, approximately  4 years
after this  event.  The  X-ray spectrum  extracted from  this latter
observation  yields an  unabsorbed  flux
$F~=~(4.46~\pm~0.05)~\times~10^{-12}~\rm{erg}~\rm{cm}^{-2}~\rm{s}^{-1}$,
using,   again,    a blackbody plus power-law model. This flux is a
factor of $\approx 6$ smaller than  that in outburst,  and is similar to
the pre-outburst flux  reported  by  Tiengo  et  al.\ (2002),  at 
$F~=~4.3~\times~10^{-12}~\rm{erg}~\rm{cm}^{-2}~\rm{s}^{-1}$   in  the
same  energy range, and is also one of the lowest fluxes reported for
this source (Oosterbroek   et  al.    1998).   We,   therefore, conclude
  that \src\  returned  to a  quiescent  state  at  the  time of  the 
2011 observation, which is further supported by the fact that no
activity has been reported since the 2007 event.  We show in
Table~\ref{obs} the details of XMM-Newton observations used in this
study.
 
Our  initial intent  was to  use both  observations for  time-resolved
spectroscopy.   However, the  high energy  particle background  showed
continuous   and   significant   variations  throughout   the   second
observation.  Eliminating  the segments  when the background  was high
and variable results in an  exposure time of $\approx 20$~ks. Together
with the  very low source count  rate in quiescence, this  prevents us
from  using   these  data   for  phase  resolved   spectral  analysis.
Therefore, we  present our results  from the timing analysis  for both
observations but  only perform phase resolved  X-ray spectral analysis
on the 2007 data.
 
\begin{table} 
\centering 
\caption{XMM-Newton Observations of the \src\ used in this study.} 
\begin{tabular}{cccc} 
\hline\hline
Observation & Date &  Observation & Exposure \\ 
Name        &      &  ID          & (ks) \\ 
\hline 
Obs1        & 2007-06-14 & 0510010601 & 48.91 \\ 
Obs2        & 2011-08-06 & 0654870101 & 96.92 \\ 
\hline 
\label{obs} 
\end{tabular} 
\end{table} 
 
Both observations were calibrated  using the Science Analysis Software
(SAS) version 11.0.0  and the latest calibration files  as of February
2012.  Barycentric correction  was applied to the  cleaned event files
using the {\it barycen} tool.   Source event files were extracted from a
circular  region  centered  on  the source  with  a  radius  of  32
arcseconds.   Similar sized  background  regions  were extracted  from
source free  regions of the  EPIC pn  detector. Given the
significantly lower effective areas of the MOS detectors and the known 
systematic difference in the flux measurements between the pn and MOS
detectors\footnote{\href{http://xmm2.esac.esa.int/docs/documents/CAL-TN
-0018.pdf}{http://xmm2.esac.esa.int/docs/documents/CAL-TN-0018.pdf}}, we
opted not to use the MOS data.  We present  below the results from the
timing and spectral analysis on these data.

\begin{figure*} 
  \centering 
\includegraphics[scale=0.5]{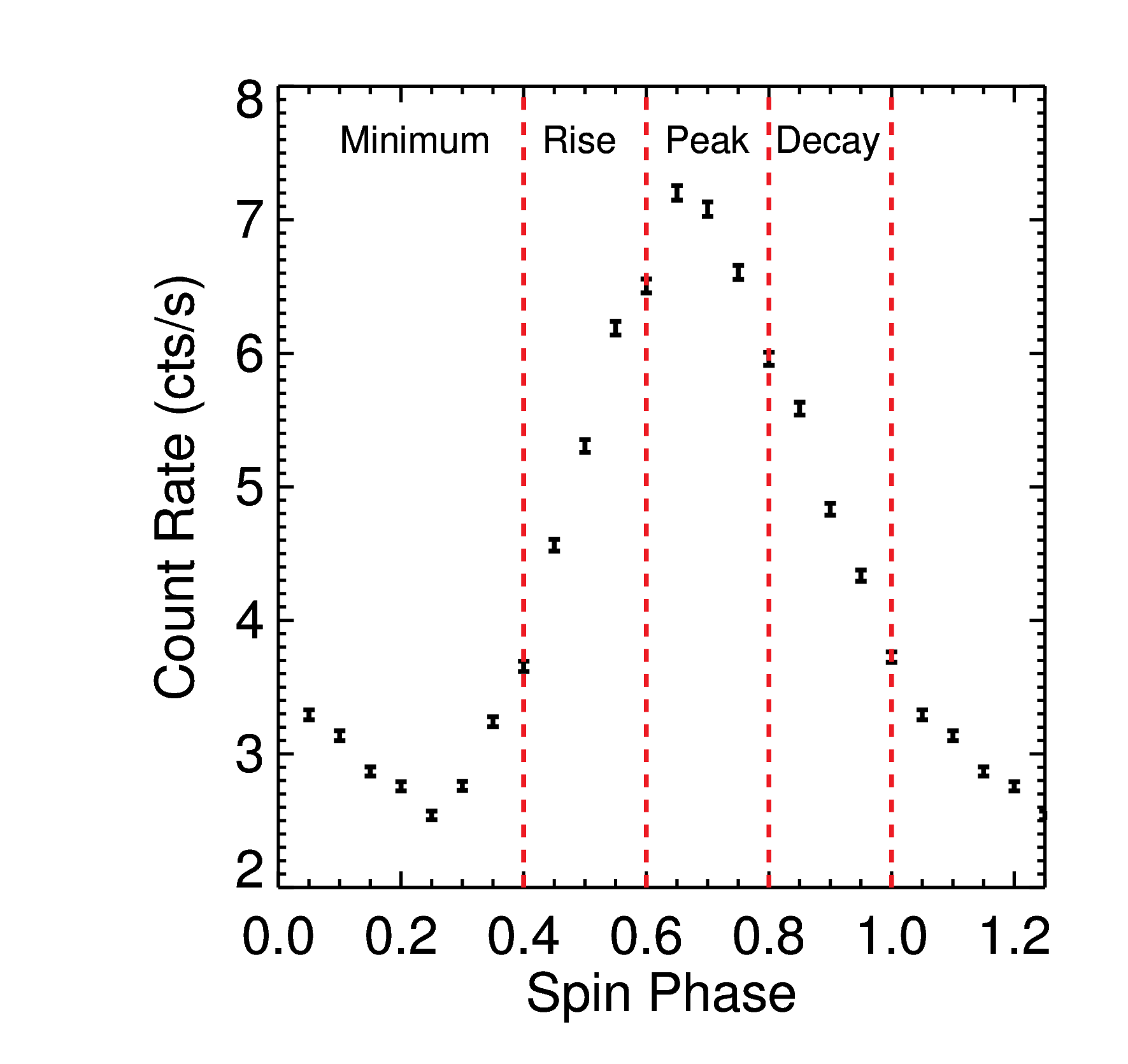} 
\caption{Pulse  profile extracted  from  the 2007  observation in  the
  2.0$-$10.0  keV  band is  shown.   Vertical  dashed lines  show  the
  intervals selected for the phase resolved spectroscopy.}
\label{ppsel} 
\end{figure*}

For  the  phase resolved  spectroscopy,  we  used  the SAS  tool  {\it
  phasecalc} to  compute the spin  phase of  each event.  In  order to
best  capture the  intrinsic structure,  while considering  count rate
limitations, we  generated phase resolved spectra  by accumulating all
events  corresponding  to  the  spin  phase  intervals  of  0.0$-$0.4,
0.4$-$0.6,  0.6$-$0.8, and  0.8$-$1.0.  The  average number  of source
counts  for each  interval  were $\approx$  58000,  54000, 65000,  and
44000,  respectively.  These  intervals represent  the minimum,  rise,
peak,   and   the  decay   of   the   pulse   profile  as   shown   in
Figure~\ref{ppsel}.
  
For  each X-ray  spectrum,  the corresponding  response and  ancillary
response files  were created using  the {\it rmfgen} and  {\it arfgen}
tools.  All X-ray  spectra were grouped to have at  least 25 counts in
each  energy  bin  and  not  to  oversample  the  instrumental  energy
resolution  more than  a factor  of three,  using the  {\it specgroup}
tool.  During the spectral fits, we added a 2\% systematic uncertainty
to  the  data,  which  represents   the  current  uncertainty  in  the
calibration of the EPIC-pn instrument\footnotemark[\value{footnote}].

\section{Results}
\label{sec_res}
\subsection{Timing Analysis} 
\label{timingsec} 
 
We used a Z$^{2}_{\rm m}$ test  (Buccheri et al. 1983) with a harmonic
number $m=1$ to  search for a periodic signal from  \src. We find very
prominent signals at frequencies of 0.15484743(3) and 0.15478524(1)~Hz
for Obs1 and Obs2, respectively.
 
\begin{figure*} 
  \centering  
\includegraphics[scale=0.5]{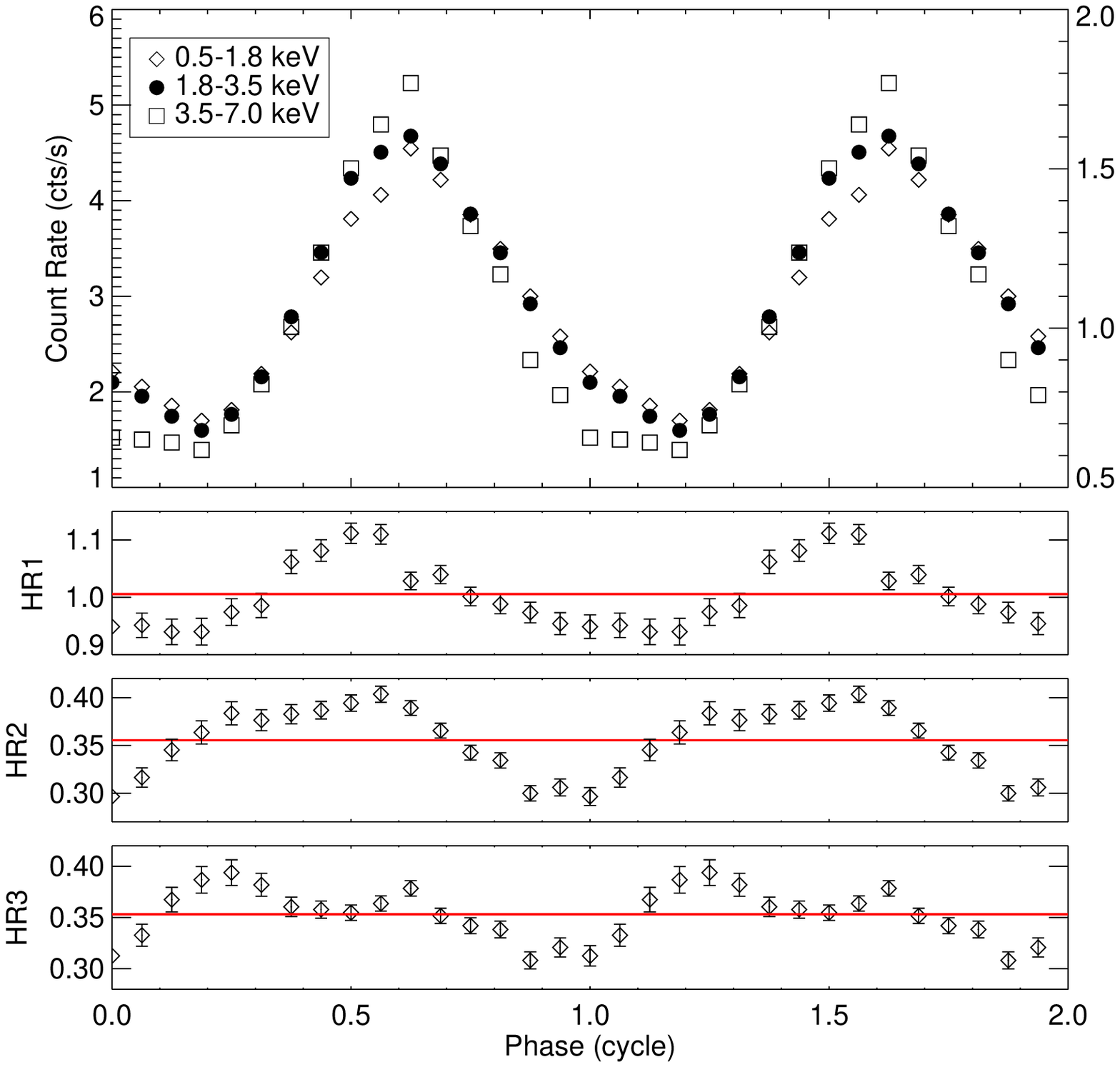} 
\caption{ {\it  Top Panel:}  Pulse profile  and its  energy dependence
  during the  observation in  2007.  The  diamond, filled  circle, and
  square symbols  show the  count rates  at 0.5$-$1.8,  1.8$-$3.5, and
  3.5$-$7.0  keV energy  ranges,  respectively.  Left  axis shows  the
  count rate  range for  0.5$-$1.8 and 1.8$-$3.5  keV bands  while the
  right axis  shows the count-rate  range for the 3.5$-$7.0  keV band.
  Lower panels show the hardness ratios for different energy ranges as
  defined in the text.  Horizontal red lines show the average hardness
  values.}
\label{ppf} 
\end{figure*}

\begin{figure*} 
  \centering 
\includegraphics[scale=0.5]{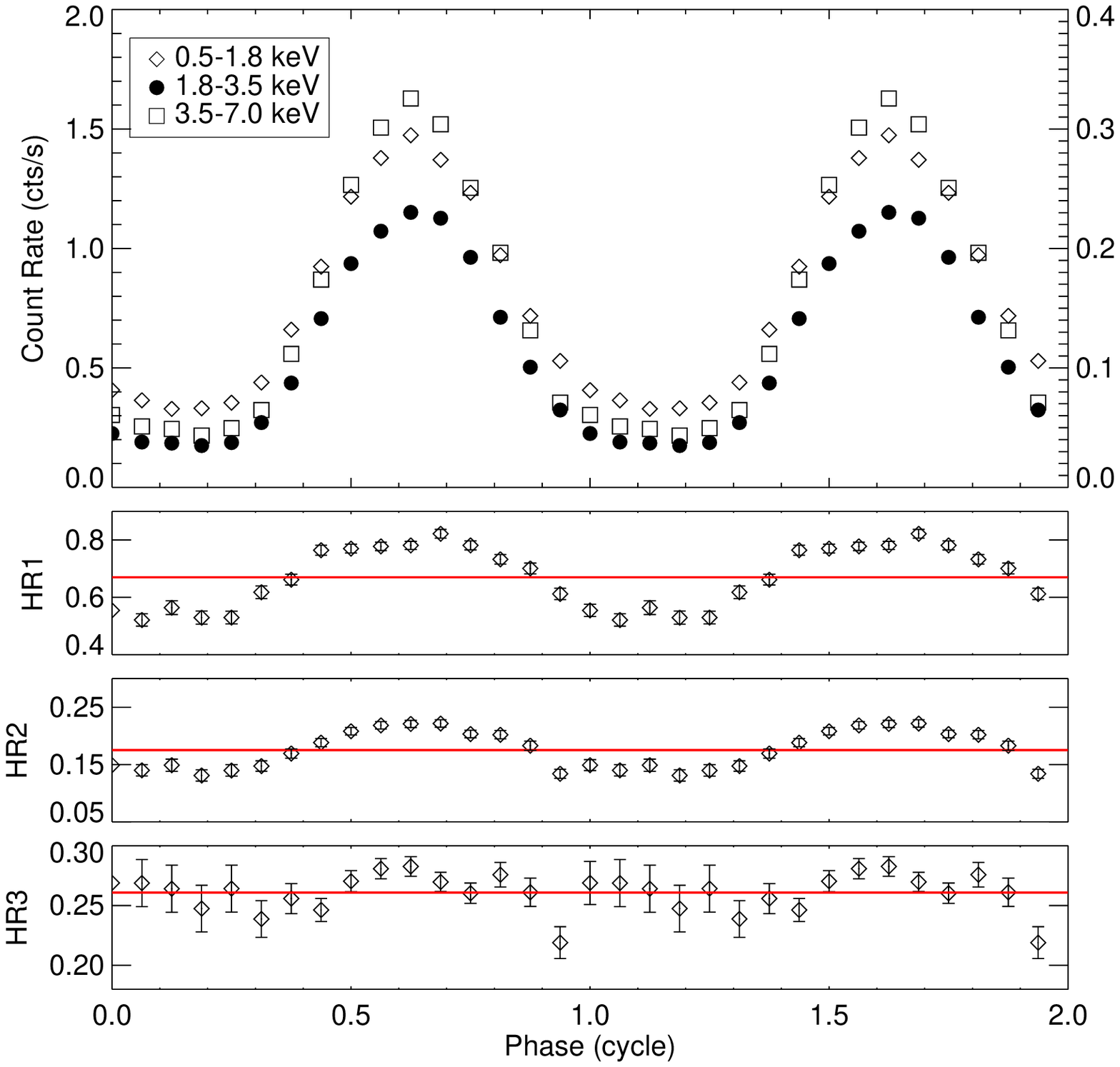} 
\caption{Pulse  profile  and its  energy  dependence  during the  2011
  observation.   Figure  labels  and  symbols   are  the  same  as  in
  Figure~\ref{ppf}.}
\label{pps} 
\end{figure*}

To investigate the energy dependence of the observed pulse profile and
of  the   pulsed  fraction,  we   generated  pulse  profiles   in  the
0.5$-$1.8~keV,  1.8$-$3.5~keV,  and   3.5$-$7.0~keV  energy  bands  by
folding  the lightcurves  in these  energy ranges  at the  frequencies
measured for  each epoch.  We chose  the above energy  ranges to
capture the characteristics  of the emission in the  softer and harder
X-ray bands as well as to keep the total number of counts in each band
as similar  as possible. We show  these profiles in the  top panels of
Figures~\ref{ppf} and  \ref{pps}.  Even though the  overall structures
of pulse profiles in the three energy bands resemble each other, there
are  discernible differences  between  them. In  particular, the  flux
(count  rate)  steadily  declines  prior  to  the  pulse  minimum  and
increases after  that in  the lower  two energy  bands of  Obs1, while
there is a plateau in the pulse minimum in all energy bands of Obs2.
 
We computed the  phase dependent hardness ratios in  order to quantify
the dependence of the pulse profiles  on photon energy.  We define the
hardness ratio HR1 as the ratio of counts per bin in the 1.8$-$3.5~keV
to those in the 0.5$-$1.8~keV band, HR2  as the ratio of counts in the
3.5$-$7.0 keV to 0.5$-$1.8~keV band, and HR3 as the ratio of counts in
the 3.5$-$7.0 keV to 1.8$-$3.5~keV band.  We show the phase dependence
of hardness ratios in the  lower three panels of Figures~\ref{ppf} and
\ref{pps} from  Obs1 and Obs2,  respectively.  We find that  HR3 shows
the  least  amount  of  variation  over the  spin  phase  during  both
observations. On  the other hand,  we find clear evidence  of spectral
variations  over  the spin  phase  below  1.8  keV:  HR1 and  HR2  are
systematically above the average hardness in the phase interval $\phi$
of 0.4$-$0.8,  and below  the mean between  $\phi \sim$  0.9$-$1.2. In
other  words, the  spectrum  during the  peak of  its  pulse phase  is
somewhat harder than that in the minimum during Obs1 and Obs2.
 
We calculated the root-mean-square (rms) pulsed fractions in the
  energy bands described above, using the Fourier based method
  described by van der Klis (1989) and applied by Woods et al. (2004)
  and Tam et al. (2008), on 1E~2259$+$586 and \src, respectively. We
also calculated it in the 2$-$10~keV band, where we can directly
compare our results with those of Tam et al.\ 2008 on data obtained
with the Chandra X-ray Observatory as \src\ went from quiescence in
2006 and to outburst in 2007.

We  list  the  pulsed  fractions in  all  energy  bands  in
Table~\ref{tab:pf}.  We  find that  the rms pulsed  fraction increases
with photon energy in both observations. In addition, we find that the
rms pulsed fraction  increased by nearly a factor two  when going from
the active to the quiescent epoch.  The pulsed fractions in all energy
bands  are anti-correlated  with  the bolometric  flux, attaining  the
higher values when the flux is  lower. The observed trend is very
  similar (but in the reverse direction) to the one reported by Tam et
  al.\ (2008), where  the flux increased from  $\sim 5 \times10^{-12}$
  to $\sim 4  \times10^{-11}$~erg~s$^{-1}$~cm$^{-2}$, while the pulsed
  fraction decreased from 0.7 to 0.3  as the source went into outburst
  (see Table~1 of Tam et al.\ 2008).

\begin{table*} 
\center 
\caption{RMS pulsed fractions for different photon energy ranges$^{*}$.} 
\begin{tabular}{ccccc} 
\hline 
\hline
 &\multicolumn{4}{c}{Pulsed Fraction}  \\ 
Observation Name  & 0.5 $-$ 1.8 keV & 1.8 $-$ 3.5 keV & 3.5 $-$ 7.0 keV & 2.0 $-$ 10.0 keV \\ 
\hline 
Obs1 & 0.311(3) & 0.350(3) & 0.362(5)   & 0.355(3) \\ 
Obs2 & 0.520(5) & 0.633(6) & 0.668(11)  & 0.648(6)\\ 
\hline 
\end{tabular} 
\footnotesize{  
\begin{flushleft}  
  $^{*}$  Values in  parenthesis  are 1$\sigma$  uncertainties in  the
  corresponding digit(s) of the measurements.
\end{flushleft} 
} 
\label{tab:pf} 
\end{table*} 


\subsection{Phase Averaged Spectral Analysis and Pulse Profile Modelling}

We first analyzed the phase-averaged X-ray spectrum extracted from the
2007 XMM-{\it Newton} observation (see Table~\ref{obs}) to determine
the average spectral parameters.  Using the phase-averaged STEMS model
(G\"uver et al.\ 2007, 2008), we obtained a very good fit
($\chi^2$/dof~=~0.93) for a surface temperature $T=0.4 \pm 0.03$~keV
and a magnetic field strength of $B =
(1.63~\pm~0.11)~\times~10^{14}$~G, where the quoted errors are
  purely statistical and do not reflect possible systematic errors or
  biases arising from the assumption of uniform temperature or
  magnetic field strength across the hot spot in the models. We also
  performed a fit to the phase averaged spectrum obtained from the
  2011 data and found best fit parameter values to be $T=0.35 \pm
  0.03$~keV and $B = (2.51~\pm~0.2)~\times~10^{14}$~G, with a
  $\chi^2$/dof~=~1.48. The statistical quality of this fit is poor
  because of the high energy particle background during this
  observation, which could not be properly accounted for. In
  particular, even after carefully removing the particle background
  using the {\it espfilt} tool, the resulting X-ray spectrum still had
  narrow residual features at around 1~keV and 2~keV regions, which
  cannot be accounted for by the continuum models. Modeling these
  features with additional Gaussian components significantly reduced
  the $\chi^2$/dof but did not affect the resulting fit
  parameters. Comparing the phase-averaged fits in the outburst and
  quiescent phases, we note that the temperatures we obtained do not
  differ significantly from one another despite the large change in
  the observed flux.

\begin{figure} 
  \centering 
\includegraphics[scale=0.45,angle=270]{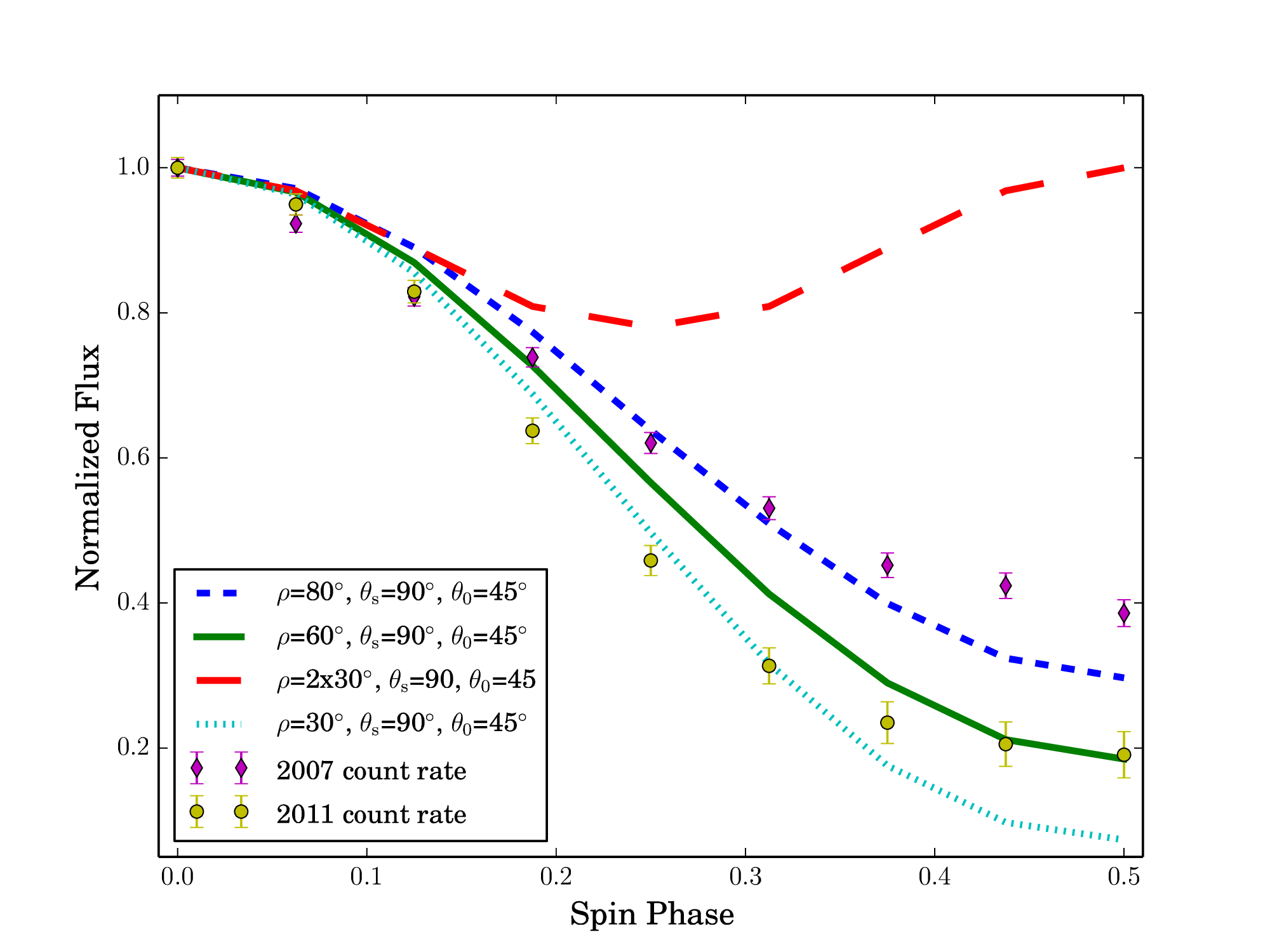} 
\caption{Pulse  profiles obtained  from  the two  observations in  the
  0.5$-$7.0 keV band are shown  with purple diamonds (Obs1) and yellow
  circles  (Obs2).   The  dotted,  solid and  dashed  lines  show  the
  theoretically   expected  pulse   profiles   with   spot  sizes   of
  30$^{\circ}$,  60$^{\circ}$, and  80$^\circ$. The  long dashed  line
  shows the case for two antipodal spots on the surface each with size
  of 30$^{\circ}$.  In  order to easily compare the  profiles, all the
  curves are normalized to the flux at  the peak and the peak phase is
  moved to phase 0.0. The plot only  extends up to phase 0.5 since the
  model assumes a symmetric emitting region.}
\label{ppss} 
\end{figure}

We used the values of the temperature and magnetic field strength
obtained from the 2007 observation to generate bolometric pulse
profile models. We compare these models to the observed pulse profiles
in order to determine the size of the hot spot in the two epochs as
well as the angles $\theta_{s}$ and $\theta_{0}$. We show in
Figure~\ref{ppss} the observed $0.5-7.0$~keV pulse profiles at the two
epochs along with a few representative models covering a range of
geometric parameters. For clarity, each profile is normalized to its
peak value and the pulse peak is moved to phase 0.0. In the model
profiles, we fix one of the angles (specifically, the spot colatitude)
at $90^\circ$ owing to the large pulse amplitudes and because the
pulsed fraction depends only on the product of the sine of the two
angles, as discussed in Section~\ref{modelsec}. We then estimate the
number and size of the hot spots as well as the observer's inclination
through a comparison with the observed pulse profiles (note that due
to the simplicity of the model geometry, we do not perform a formal
fit).  It is immediately clear from this figure that a configuration
with two antipodal hot spots (shown in the long-dashed line) has the
wrong number of peaks and cannot even qualitatively describe the
observed profiles.  In contrast, a single hot spot, with a colatitude
of $90^\circ$ and an observer's inclination of $45^\circ$ results in a
reasonable match to the observed pulse amplitude (see the discussion
below for the pulse minimum). 

The spot sizes we obtain for two epochs are significantly different:
the pulse profile extracted from the 2007 observation can be
reproduced with a spot size $\rho \approx 80^\circ$, while the one in
2011 can be reproduced with a spot size of $\rho \approx 30^\circ$.
The resulting difference implies an $ \approx 85\% $ decrease in the
emitting area from outburst back into quiescence.  This is in
agreement with expectations from a model where the outburst is caused
by thermal changes on the neutron star surface, where the shrinking
hot spot size during the decline simultaneously accounts for the
observed increase in the pulsed fraction and the decrease in the X-ray
flux. We use these results to fix the three geometric parameters in
the subsequent analysis of the phase-resolved spectra in
Section~\ref{spec_res}.

\begin{figure} 
  \centering 
\includegraphics[scale=0.45,angle=270]{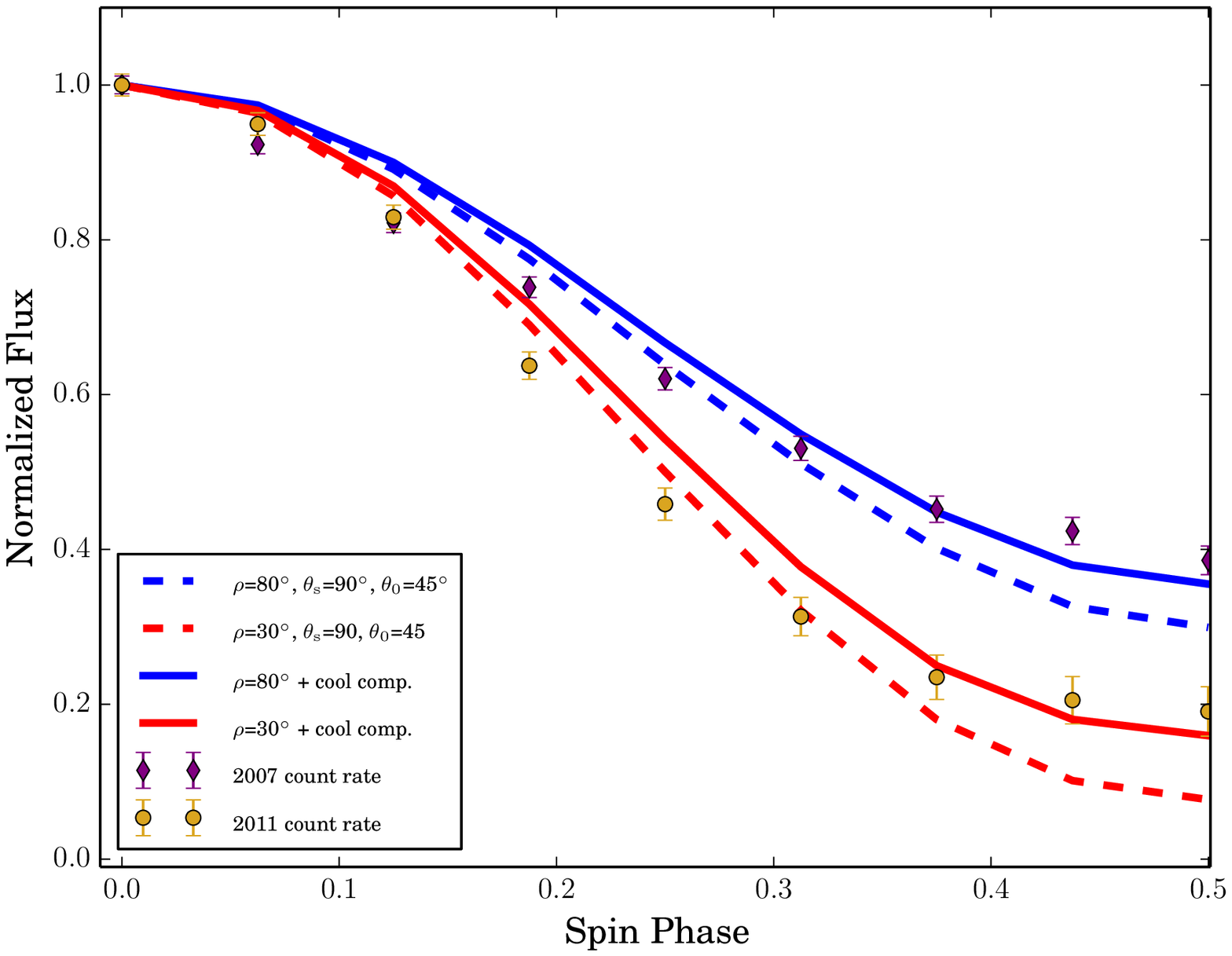} 
\caption{ Model  pulse profiles obtained for  the $80^\circ$ (blue
    solid line) and  $30^\circ$ (red solid line)  hot spots described
    in Figure 5  but with the addition of an  antipodal cool component
    that has a temperature of 0.2~keV and 0.15~keV, respectively.  For
    comparison, the  dashed lines  show the models  with only  one hot
    spot, with parameters as in Figure 5. Observed 0.5$-$7.0~keV pulse
    profiles are shown with purple  diamonds (Obs1) and yellow circles
    (Obs2).}
\label{ppss2} 
\end{figure}

While the model profiles describe the observed pulse shapes well at
the peak and decay pulse phases, they underpredict the flux at pulse
minima. This behavior is most likely related to our assumption that
the stellar surface outside of a localized hot region is dark and does
not contribute at all to the emission.  In reality, it is highly
likely that the rest of the surface, which is cooler and covers a
larger area, also has a small contribution to the observed emission
and may become most visible at the pulse minimum, when the hot spot is
half a spin phase away from the observer's line-of-sight.  To test
this interpretation, we calculated models where we added a large
antipodal cool component, but kept all the parameters describing the
magnetic field strength, the size, and the geometry of the hot spot
the same. We show in Figure~\ref{ppss2} the resulting pulse profiles
that include a cool spot with an angular size $\rho=80^\circ$ and a
temperature of T~=~0.2~keV and 0.15~keV, respectively, for the 2007
and 2011 observations.  This additional component indeed helps provide
a better match to the flux at the pulse minimum. Because we are
interested in the characteristics of the hot, active region and the
changes to these characteristics between the quiescent and outburst
epochs, we do not consider the contribution from this cool component
in the rest of the analyses in this paper.

\subsection{Phase Resolved Spectral Analysis} 
\label{spec_res} 
 
We showed in Section~\ref{timingsec} that the flux observed from
\src\ depends significantly on the pulse phase. Equivalently, the
observed pulse shapes show a strong energy dependence and a
significant change between the two observations. To understand in more
detail the causes of the phase dependent flux modulations and to
assess if these modulations can be used to constrain the magnetic
field strength and temperature distribution, we carry out in this
section phase resolved X-ray spectroscopy using the model described
Section~2.
  
\begin{figure} 
  \centering 
\includegraphics[scale=0.35]{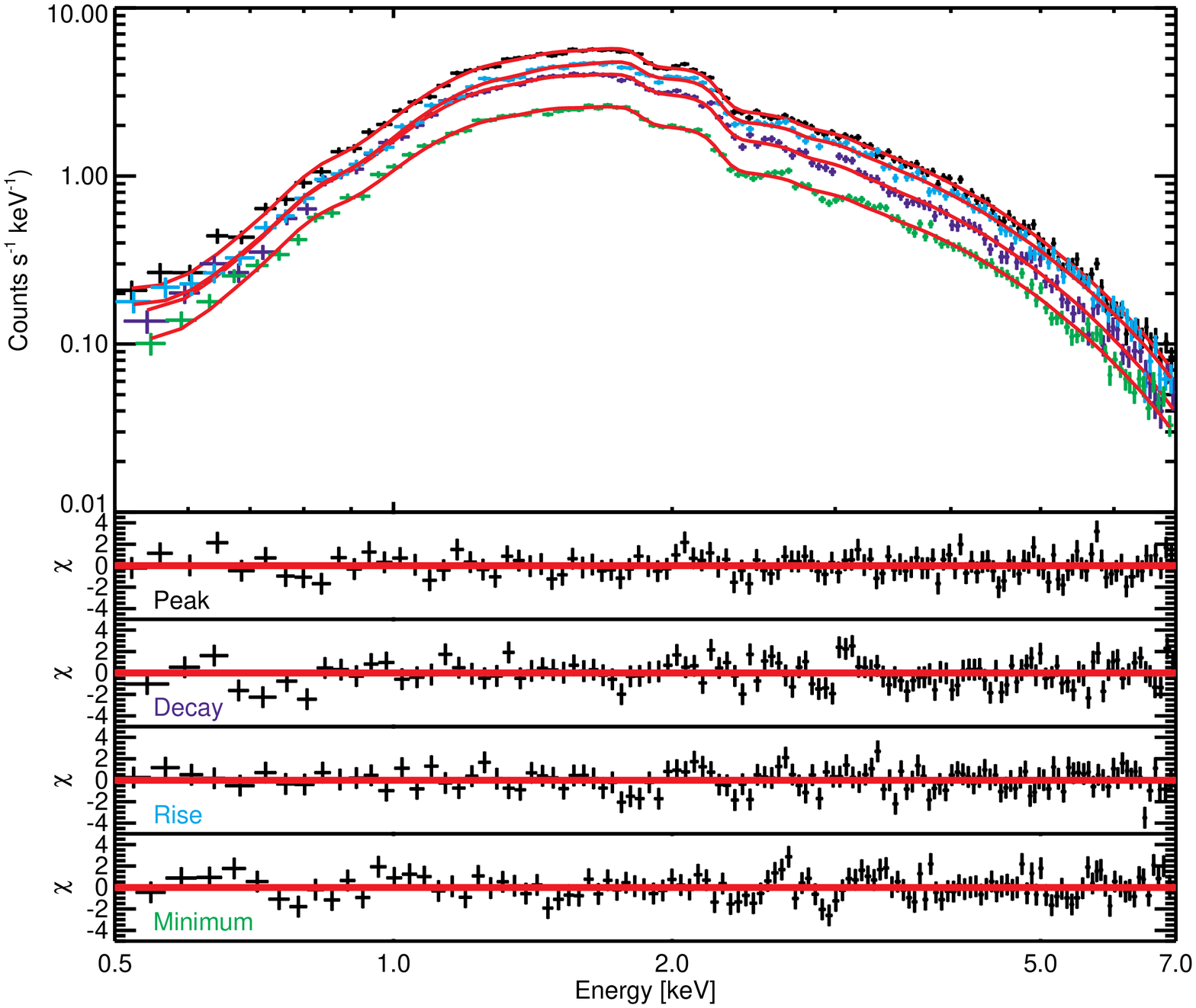} 
\caption{Upper panel  shows the best  fit models  to each data  set at
  different spin phase intervals. Lower panels show the residuals from
  the best fit models for individual spin phases.}
\label{resfig} 
\end{figure}

\begin{figure} 
  \centering 
\includegraphics[scale=0.35]{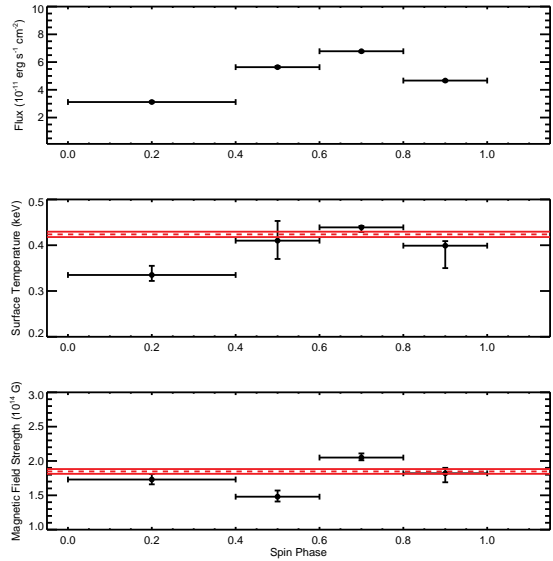} 
\caption{ Evolution of  the 0.5$-$7.0~keV  unabsorbed flux,  the
  surface  temperature and  the  magnetic field  strength  of the  hot
  region on the surface as a  function of neutron star spin phase. Red
  dashed  and solid  lines show  the  error weighted  averages of  the
  individual best fit values and their 1$-\sigma$ uncertainties.}
\label{phase_ev} 
\end{figure}

We fit the spectra obtained  at different pulse phases simultaneously,
linking  only  the  hydrogen  column density  and  the  magnetospheric
parameters  $\beta$ and  $\tau$  between spin  phase intervals,  since
these parameters are not expected  to vary throughout the neutron star
spin.  We  obtained a  fit with  $\chi^{2}$/dof =   1.07  for 477
degrees of freedom.  We list the  parameters obtained in each phase in
Table~\ref{freeres}  and show  the  best-fit models  and residuals  in
Figure~\ref{resfig}. We  also plot  in Figure~\ref{phase_ev}  the spin
phase evolution of the best fit parameters.
 
Overall, we find that a model consisting of a rotating hot spot with a
size of 80$^{\circ}$ and a temperature of $\approx$~0.4 keV provides a
good fit to the data at all spin phases  (but see the next section
  for  a discussion  of the  pulse  minimum).  We  obtain an  average
magnetic  field  strength  of   $1.8\times10^{14}$~G.   The  best  fit
magnetic field  strength shows a 13\%  rms variation as a  function of
spin phase, which is only at  a 2$-\sigma$ level given the statistical
uncertainties.  Variations  of this  magnitude are expected  given the
simplifications in the model, such as the assumption that the magnetic
field  is uniform  throughout the  emitting  region.  We  also find  a
best-fit  magnetospheric  optical   depth  $\tau=4.86\pm0.17$  and  an
average  particle  velocity  $\beta=  0.300\pm0.005$.   Rea  et
al.\ (2008)  modeled the same  observation using a  resonant cyclotron
scattering  model,  where a  blackbody  spectrum  is modified  through
resonant  cyclotron   scattering  by  charges  in   the  neutron  star
magnetosphere.   They   reported  the  magnetospheric   parameters  as
$\tau=4.7\pm0.2$ and $\beta=0.29\pm0.05$, in  very good agreement with
the results we found here.

\begin{deluxetable}{cccccccc}
\tablecolumns{8}
\tablewidth{0pc}
\tablecaption{Results of the phased  resolved spectroscopic analysis of the 
  2007 observation.}
\tablehead{
\colhead{Phase} & \colhead{N$_{\rm H}$$^{1}$} & \colhead{kT} & \colhead{B}
 & \colhead{$\beta^{1}$} &  \colhead{$\tau^{1}$} & \colhead{Flux$^{2}$} & $\chi^{2}_{\nu}$\\
 \colhead{}    &  \colhead{(10$^{22}\rm{cm}^{-2}$)}  & 
 \colhead{(keV)}   & \colhead{(10$^{14}$ G)} & &  & \colhead{(dof)}}
\startdata
Peak &  0.80$_{-0.02}^{+0.01}$ & 0.439$^{+0.003}_{-0.010}$ & 2.05$^{+0.06}_{-0.04}$ & 0.300$\pm$0.005  & 4.86$\pm$0.17 & 6.78$_{-0.03}^{+0.07}$ & 1.0708 (477)\\   
Decay &  --           & 0.399$^{+0.003}_{-0.049}$ & 1.80$^{+0.08}_{-0.13}$ & -- &  --                                               & 4.67$^{+0.06}_{-0.03}$ & -- \\ 
Rise & --             & 0.410$^{+0.043}_{-0.040}$ & 1.48$^{+0.09}_{-0.07}$ &      --                   & --                         & 5.63$^{+0.07}_{-0.09}$ & -- \\ 
Minimum &  --         & 0.335$^{+0.020}_{-0.013}$ & 1.73$^{+0.08}_{-0.07}$ & -- & --                                                & 3.12$_{-0.05}^{+0.04}$ & -- \\ 
\enddata
\footnotesize{\begin{flushleft} $^{1}$  Best-fit values  were obtained
    by linking the parameters over all spin phase intervals. \\
 $^{2}$ Unabsorbed 0.5$-$7.0~keV flux in units of 10$^{-11}$ erg s$^{-1}$ cm$^{-2}$.
\end{flushleft}} 
\label{freeres} 
\end{deluxetable}

\section{Discussion} 
\label{disc} 

In this paper, we investigated the pulse profile and pulse phase
resolved X-ray spectra of \src\ to map its surface temperature and
measure its magnetic field in outburst and in quiescence.  We fit two
epochs of XMM data using a spectral model that takes into account the
relevant physical and geometrical effects that includes: {\it (i)} a
strongly magnetized fully ionized hydrogen atmosphere on the neutron
star surface characterized by an effective temperature $T$ and a
magnetic field strength $B$ and yields the energy distribution and
beaming of the emitted photons, {\it (ii)} the scattering of these
surface photons in the magnetosphere due to the existence of a dense
charged particle cloud with a total scattering optical depth $\tau$
and an average electron velocity $\beta$, {\it (iii)} the effects of
gravitational lensing and redshift for a slowly rotating neutron star
to determine the observables at infinity.  These calculations also
account for the number and size of the hot spots on the surface, the
inclination of the observer's line-of-sight and the colatitude of the
hot spot with respect to the stellar spin axis.
 
The high pulsed fraction of \src\ allows us to constrain the number of
the hot  spots on the  surface to one, the  colatitude of the  spot to
$\theta_s  \approx  90^{\circ}$,  and the  observer's  inclination  to
$\theta_0  \approx 45^{\circ}$.    Note that,  because these  two
  parameters are degenerate as discussed in Section~2, a configuration
  with $\theta_s \approx 45^{\circ}$ and $\theta_0 \approx 90^{\circ}$
  is also equally likely.  Furthermore, using the spectral parameters
found from the  phase averaged spectrum obtained in  2007 and assuming
that  the angle  between  the rotation  axis and  the  observer to  be
$\approx 45^{\circ}$,  we show  that the pulse profiles  are best
  described by models where the angular sizes of the  hot active
region  are  80$^{\circ}$ in  2007  and  30$^{\circ}$ in  2011.   
  Therefore, in 2007,  the angular diameter of the  emitting region is
  larger  than that  in 2011  by  a factor  of $\approx  6$, which  is
  identical  to  the observed  change  in  the  flux between  the  two
  epochs. This suggests that after the timing glitch observed in 2007,
  the  size of  the  hot spot  on the  neutron  star surface  changed,
  increasing to about $80^\circ$ and  causing the observed increase in
  the  X-ray  flux. By  2011,  it  returned  close to  its  pre-burst,
  quiescent level of $30^\circ$. 

If a  single rotating hot  spot described by one  temperature captures
the  physical conditions  relevant for  the surface  of \src,  then we
expect  to   only  see   insignificant  differences  in   the  surface
temperatures  obtained  from  the   phase-resolved  spectra  that  are
consistent within statistical uncertainties.   For the peak, rise, and
decay phases, this is indeed what we obtain: the standard deviation of
the  best fit  temperature for  these phases  is equal  to the  formal
uncertainties     of     the      individual     measurements     (see
Figure~\ref{phase_ev}),  indicating that  a single  rotating hot  spot
with an  angular size of  $\approx 80^{\circ}$ dominates  the observed
flux during  these spin  phases.   This  model, however,  does not
adequately account  for the flux  in the  phase minimum as  shown in
Figure~\ref{ppss}, where the rest of  the surface, albeit being much
cooler,  is expected  to contribute  to the  emission. We  showed in
Section~4.2 and Figure~\ref{ppss2} that adding the contribution from
such a cool component that  is $180^\circ$ out-of-phase with the hot
spot indeed  leads to  a better  description of  the profile  at the
pulse  minimum.  This  component  will naturally  introduce a  small
modification to  the spectral parameters  obtained from the  fits at the
pulse  minimum and may  account for the discrepancy  between the
temperatures  at  the phase  minimum  and  the other  three  phases.
Testing this hypothesis is  beyond the computational capabilities of our
current setup. 

We compare the magnetic field strengths we derived here with other
independently obtained estimates of the field in this source.  We
obtained an average magnetic field strength of $1.8\times10^{14}$~G,
covering a range from 1.48 to $2.05 \times 10^{14}$~G at different
pulse phases, which is a scatter at the $2\sigma$-level given the
statistical uncertainties in each of these measurements. In an earlier
study where we applied the STEMS model to the phase-averaged spectrum
obtained from the 2003 XMM-Newton observation of this source, we
reported an average field strength of $(2.26\pm0.05)\times10^{14}$~G
(\"Ozel, G\"uver, \& G\"o\u{g}\"u\c{s} 2008), which is similar to the
present findings. The small difference is caused by the fact that
general relativistic transport was applied to the average spectrum as
a whole in this earlier study rather than to spectra at each pulse
phase; this introduces a difference in the spectrum observed at
infinity when the emerging radiation is highly beamed. There are
  naturally other sources of systematic uncertainty or bias arising
  from our assumption of constant magnetic field strength and
  direction throughout the polar cap and in the magnetosphere. In the
  absence of first-principles calculations of the magnetic field
  topology and the strength of the currents in the magnetosphere, it
  is difficult to predict the direction of this bias. However, the
  close similarity between the magnetic fields inferred from
  pulse-phase averaged spectra and those of the individual pulse
  phases suggests that the geometric effects such as those we
  neglected here do not change the results substantially. An
estimate of the dipole magnetic field strength can also be obtained
from the rate of spindown in pulsars.  \src\ has a spin period of
6.452 s, with a period derivative that shows large excursions in the
range $\dot P = (0.86 - 3.81) \times 10^{-11} \rm{ss^{-1}}$ (Kaspi et
al.\ 2001).  Using the standard magnetic dipole radiation formula, the
range of magnetic field strengths implied by these measurements is
$B_{\rm dip} = (2.4 - 4) \times 10^{14}$~G, where we assumed a neutron
star radius of $R=10$~km and a moment of inertia of
$I=10^45$~g~cm$^{2}$.  This shows a reasonable agreement between the
field strengths obtained from spectral and timing methods.

Finally, we  combine the  surface areas we  obtain from  pulse profile
modeling  and phase  resolved  spectroscopy  with previously  reported
values  of  the  pulsed  fraction  of \src\  at  different  epochs  to
reconstruct  a  physical  picture  of its  recent  activity.   Tam  et
al.\   (2008)   reported  a   $2-10$~keV   RMS   pulsed  fraction   of
$\approx$~70\% for \src, using several Chandra observations throughout
2006 when the source showed no activity. In March 2007, \src\ showed a
pulsed flux increase,  which was followed by a spin-up  glitch (Dib et
al.\  2009).  Subsequent  observations with  imaging X-ray  telescopes
revealed that  the total flux  increased by  a factor of  $\approx 7$,
while the pulsed fraction  decreased significantly from $\approx$~70\%
to $\approx$~30\%  (Tam et al.\ 2008).   The observed anti-correlation
between the  total flux and  the pulsed  fraction was attributed  to a
growing hot  spot on the neutron  star surface (Tam et  al.\ 2008; see
\"Ozel \&  G\"uver 2007).  Our  findings support this  hypothesis.  We
measured a pulsed fraction of 35\% in the 2$-$10 keV band as well as a
spot size of  $\approx 80^{\circ}$ in the 2007  observation, which was
performed approximately  3 months after  the onset of the   glitch
  event  (Dib   et  al.\  2009).    The  pulsed  fraction   is  still
significantly lower than that before  the flux increase but shows that
the neutron star was already recovering after the 2007 event. In 2011,
we observe that  the pulsed fraction returned to  $\approx 65$\% level
while  the size  of the  hot spot  decreased to  $\approx 30^{\circ}$.
These  results  suggest that  the  emitting  area grows  significantly
during flares  and glitches  but recovers back  to its  persistent and
significantly smaller size within a few year timescale.

\section*{Acknowledgments}
We  thank the  anonymous  referee for  his/her  detailed comments  and
suggestions.   TG   acknowledges  support  from  the   Scientific  and
Technological Research Council of Turkey (T\"UB\.ITAK B\.IDEB) through
a fellowship  programme.  F.\"O. gratefully acknowledges  support from
NSF grant AST-1108753.

\label{lastpage} 
\end{document}